\documentclass[11pt,a4paper]{article}
\usepackage{jheppub}
\usepackage{amsmath,amssymb}
\usepackage{slashed}
\usepackage{graphicx}%
\usepackage{subfigure}
\usepackage{mathptmx}
\usepackage{url}
\usepackage{axodraw4j}
\newcommand{\beq}{\begin{equation}}
\newcommand{\eeq}{\end{equation}}

\newcommand{\be}{\begin{eqnarray}}
\newcommand{\ee}{\end{eqnarray}}
\long\def\hidestart#1\hideend{}
\title
{Many avatars of the Wilson fermion: A perturbative analysis}
\author{Abhishek Chowdhury$^{a}$,}
\author{A. Harindranath$^{a}$,}
\author{Jyotirmoy Maiti$^{b}$ and}
\author{Santanu Mondal$^{a}$}

\affiliation{$^{a}$Theory Division, Saha Institute of Nuclear Physics \\
 1/AF Bidhan Nagar, Kolkata 700064, India}

\affiliation{$^{b}$Department of Physics, Barasat Government College,\\
10 KNC Road, Barasat, Kolkata 700124, India}

\emailAdd{abhishek.chowdhury@saha.ac.in}
\emailAdd{a.harindranath@saha.ac.in}
\emailAdd{jyotirmoy.maiti@gmail.com}
\emailAdd{santanu.mondal@saha.ac.in}
\date{December 20, 2012}
\abstract { We explore different branches of the fermion doublers with Wilson fermion 
in perturbation theory, in the context of additive mass 
renormalization and chiral 
anomaly, and  show that by appropriately averaging over suitably
chosen branches one can reduce cut-off artifacts. Comparing the central 
branch with
all other branches, we find that the central branch, among all the avatars 
of the Wilson fermion, is the most suitable candidate
for exploring near conformal lattice field theories.}
\begin{document}
\maketitle

\section{Introduction}
The most reliable techniques to investigate various non-perturbative aspects 
of quantum field theories are provided by lattice methods. Putting 
fermion on a lattice, however, has turned out to be highly non-trivial because
of the notorious doubling problem. Naive discretization of the Dirac action 
leads to 16 solutions (called doublers) in the four dimensional theory.
Among the various solutions suggested to 
cure this problem, Wilson fermions \cite{wilson} are conceptually the 
simplest and straightforward to implement. In the Wilson approach a 
dimension five operator is added to the action thereby sending the masses of 
the extra fifteen fermions to the order of the cutoff. Thus the extra 
fermions decouple in the continuum limit.    
It is well-known that the Wilson 
formulation of lattice gauge theory preserves discrete symmetries of the continuum 
formulation which simplifies the construction of lattice operators that 
correspond to the observables in the continuum theory. The Wilson 
term in the action, which is  introduced to remove 
the doublers, breaks chiral symmetry in accordance with 
the well-known Nielson-Ninomiya no-go theorem \cite{nn}.
Wilson term reproduces the correct axial anomaly 
\cite{ks,kerler,rothe} eventhough it leads to additive renormalization for 
the fermion mass. 

The sixteen doublers are classified into five branches. 
Almost all of the studies so far, both analytical and numerical, 
have focused on the so-called first (physical) branch. However, very recently,
occurrence of an enhanced symmetry has been discovered in the central branch 
\cite{creutz, kimura, misumi} when the on-site terms (mass term and that 
from the Wilson term) are absent in the action. The enhanced symmetry prohibits
 additive
renormalization through radiative corrections. Since in this case, the central 
branch yields six massless fermions, as suggested by ref. \cite{kimura}, 
an alternative way to simulate twelve 
flavour non-abelian gauge theories emerges. Such theories are of interest in the context of
beyond standard model physics (for recent reviews, 
see \cite{iwasaki, giedt, neil, deldebbio}).     
  
In this work, by introducing a generalized Wilson term containing a branch 
selector index, we investigate the additive fermion mass shift and chiral 
anomaly to ${\cal O}(g^2)$ in lattice perturbation theory for all the 
branches of the fermion doublers. 
\section{Preliminaries}
We denote the generalized Wilson fermion action by 
\begin{eqnarray}
 S_F[\psi,{\overline\psi},U](i_B) & = & a^4~\sum_{x,y}{\overline \psi}_{x}
M_{xy}(i_B)\psi_{y} = 
  a^4~\sum_{x,y}{\overline \psi}_{x} 
\left [ \gamma_\mu D_\mu + W(i_B) +m \right ]_{xy}\psi_y~~  {\rm with} \\
{[D_\mu]}_{xy} &=&  \frac{1}{2a}~\left [U_{x,\mu }~\delta_{x+\mu,y}  -
U^\dagger_{x-\mu, \mu}~\delta_{x-\mu,y} \right ]~~~~~{\rm and } \\
W_{xy}(i_B) &=&  \frac{r}{2a}~\sum_\mu \left [ 2 (1 - \frac{1}{2} i_B)~
\delta_{x,y} -
U_{x,\mu }~\delta_{x+\mu,y}  - 
U^\dagger_{x-\mu, \mu}~\delta_{x-\mu,y} \right ]~.
\end{eqnarray}
The factor $i_{\rm B}$ is the branch selector 
index which takes the values 0,
1, 2, 3, and 4 for first, second, third (central), fourth and fifth 
branch of the doubler respectively.
We consider the transformations \cite{kimura}
\begin{eqnarray}
\psi_x \rightarrow \psi^{\prime}_x = e^{i\theta(-1)^{x_1+x_2+x_3+x_4}}\psi_x,~~
{\overline\psi}_x \rightarrow {\overline\psi}^{\prime}_x 
= e^{i\theta(-1)^{x_1+x_2+x_3+x_4}}{\overline\psi}_x~~.
\label{symmetry}
\end{eqnarray}
The action is invariant under these transformations but for the local terms.
For $m=0$ and $i_B=2$ (massless limit of the central branch), the action thus
possesses this additional symmetry which prevents additive renormalization of 
the fermion mass through radiative corrections. 


\section{Additive Renormalization in Fermion Self Energy}
In this section we calculate the additive shift to ${\cal O}~(g^2)$ in the fermion mass
(for $am=0$) using lattice perturbation theory \cite{lpt}.

\begin{figure}[h]
\begin{center}
\fcolorbox{white}{white}{
  \begin{picture}(148,109) (303,-219)
    \SetWidth{1.0}  
    \SetColor{Black}
    \Line[arrow,arrowpos=0.5,arrowlength=5,arrowwidth=2,
arrowinset=0.2](304,-219)(432,-219)




    \GluonArc(368,-179)(32,-180,180){7.5}{16}
    \Text(320,-217)[lb]{\Large{\Black{p}}}
    \Text(416,-217)[lb]{\Large{\Black{p}}}
    \Text(368,-131)[lb]{\Large{\Black{k}}}
  \end{picture}
}
\end{center}
\caption {Tadpole diagram }
\label{tadpole}
\end{figure}
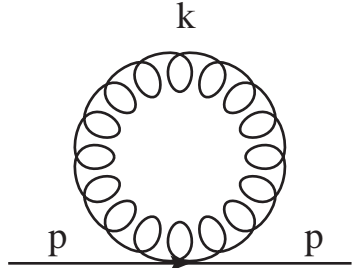

The additive shift in fermion mass due to the tadpole diagram (figure
\ref{tadpole}) is
\begin{eqnarray}
\delta m ~ = ~  -\frac{r}{a}~\frac{1}{2} g^2~ C_F~ \sum_\mu \cos(ap_\mu) ~ Z_0
\end{eqnarray}
 with
$Z_0 = \int \frac{d^4k}{(2 \pi)^4}~ \Bigg(4 \sum_\lambda 
\sin^2\Big(\frac{ak_\lambda}{2}\Big)\Bigg)^{-1}~$ and $C_F=\frac{N^2-1}{2N}$ for SU(N).
 
Results for different branches are as follows.\\
First branch: $ap_\mu = (0,~0,~0,~0)  \rightarrow \delta m =
-2 \frac{r}{a} g^2 ~ C_F ~ Z_0~ $. \\
Second branch: $ap_\mu = (\pi,~0,~0,~0)~ {\rm or ~any ~of~ the ~ other 
~ three ~ permutations}  \rightarrow \delta m =
-1 \frac{r}{a} g^2 ~ C_F ~ Z_0~ $. \\ 
Third (central) branch: $ap_\mu = (\pi,~\pi,~0,~0)~ {\rm or ~any ~of~ the ~ 
other ~ five ~ permutations}  \rightarrow \delta m ~=~0.$ \\
Fourth branch: $ap_\mu = (\pi,~\pi,~\pi,~0)~ {\rm or ~any ~of~ the ~ other 
~ three ~ permutations}  \rightarrow \delta m =
+1 \frac{r}{a} g^2 ~ C_F ~ Z_0~ $. \\
Fifth branch: $ap_\mu = (\pi,~\pi,~\pi,~\pi)  \rightarrow \delta m =
+2 \frac{r}{a} g^2 ~ C_F ~ Z_0~ $. \\
Next consider the additive mass shift in fermion  mass due to sunset diagram.  
The gauge boson propagator in Feynman gauge is given by
\begin{eqnarray}
G^{ab}_{\mu\nu}=\delta_{\mu\nu}\delta^{ab}\Big\{\frac{4}{a^2}
\sum_{\lambda}\sin^2\frac{a(p-k)_\lambda}{2}\Big\}^{-1}~=~
\delta_{\mu\nu}\delta^{ab}~ \Big\{(1/a^2)~{\cal W}_{p,k}\Big\}^{-1}~,
\end{eqnarray}
whereas the fermion propagator has the form
\begin{eqnarray}
S^{lm}(k,i_B)=\delta^{lm}\Bigg\{\sum_{\mu}i\gamma_{\mu}
\frac{\sin(k_{\mu}a)}{a}+\frac{r}{a}{\cal M}_k(i_B)
\Bigg\}^{-1}
\end{eqnarray}
with
\begin{eqnarray}
 {\cal M}_k(i_B) = 
\sum_\mu \Bigg [ \Big(1 - \frac{1}{2} i_B \Big) ~ - ~ \cos (k_\mu a) \Bigg ] 
\end{eqnarray}
and the fermion-gauge boson vertex is 
\begin{eqnarray}
(V^{a})^{mn}_{\rho}(k,p)= -g (T^a)^{mn}\Big\{i\gamma_{\rho}\cos\frac{a(k+p)_{\rho}}{2}
+r\sin\frac{a(k+p)_{\rho}}{2}\Big\}~.
\end{eqnarray}
Then the fermion self energy from sunset diagram can be evaluated as
\begin{eqnarray}
\Sigma= \int \frac{d^4k}{(2 \pi)^4}~\sum_{\rho}G^{ab}_{\rho\rho}(p-k)
(V^b)^{lm}_{\rho}(k,p)S^{mn}(k)
(V^a)^{nl}_{\rho}(p,k)~.
\end{eqnarray} 
The additive mass shift arising from the fermion self energy (sunset) can be written as 
\begin{eqnarray}
\delta m =\frac{r}{a} g^2 C_F \int \frac{d^4k}{(2 \pi)^4} \frac {N_r}{D_r} 
\end{eqnarray}
where $D_r~=~{\cal W}_{p,k}\left(\Gamma^2+r^2{\cal M}_k^2(i_B)\right) $ with 
 $\Gamma^2~=~\sum_\lambda\sin^2 (ak_\lambda) $.
 We introduce $\Gamma_\lambda~=~\sin (ak_\lambda) $, $ {S}_\rho~=~ 
\sin (\frac{ak_\rho}{2})$ and 
$ {C}_\rho~=~ \cos (\frac{ak_\rho}{2})$. The expressions for $N_r$ and $D_r$ for
different branches are given below. 
\begin{figure}[h]
\begin{center}
\fcolorbox{white}{white}{
\begin{picture}(178,161) (271,-170)
    \SetWidth{1.0}
    \SetColor{Black}
    \Line[arrow,arrowpos=0.5,arrowlength=5,
arrowwidth=2,arrowinset=0.2](272,-159)(448,-159)


    \GluonArc[clock](360,-118)(57.28,-134.293,-405.707){7.5}{22}
   \Text(290,-157)[lb]{\Large{\Black{p}}}
    \Text(360,-157)[lb]{\Large{\Black{k}}}
  \Text(430,-157)[lb]{\Large{\Black{p}}}  
   \Text(357,-90)[lb]{\Large{\Black{p-k}}}
    %
  \end{picture}
}
\caption{Sunset diagram}
\end{center}
\label{sunset}
\end{figure}
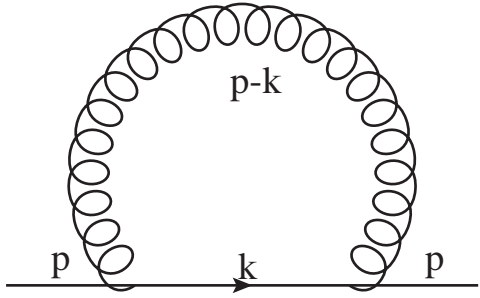

\noindent First branch: $ap_\mu~=~(0,~0,~0,~0)$. 
\begin{eqnarray}
N_r &=& \sum_{\rho=1}^4 \Big[{\cal M}_k(i_B=0)
(r^2 S_\rho^2-C_\rho^2)+ \Gamma_\rho^2\Big]~,\\
D_r &=& {\cal W}_{p,k}\Big[\Gamma^2+r^2{\cal M}_k^2(i_B=0)\Big]~.
\end{eqnarray}
Second branch: $ap_\mu~=~(\pi,~0,~0,~0)$ or 
three other permutations. Explicitly 
for $ap_\mu~=~(\pi,~0,~0,~0)$  
\begin{eqnarray}
N_r &=&- \Big[{\cal M}_k(i_B=1)( S_\rho^2-r^2 C_\rho^2)+ 
\Gamma_\rho^2\Big]_{\rho=1}
 + \sum_{\rho=2}^4 \Big[{\cal M}_k(i_B=1)(r^2 S_\rho^2- 
C_\rho^2)+ \Gamma_\rho^2\Big]~, \\
D_r &=& {\cal W}_{p,k}\Big[\Gamma^2+r^2{\cal M}_k^2(i_B=1)\Big]~.
\end{eqnarray}
Third (central) branch: $ap_\mu~=~(\pi,~\pi,~0,~0)$ or five other 
permutations. \\
Explicitly 
for $ap_\mu~=~(\pi,~\pi,~0,~0)$  
\begin{eqnarray}
N_r &=&- \sum_{\rho=1}^2 \Big[{\cal M}_k(i_B=2)( S_\rho^2-r^2 C_\rho^2)
+ \Gamma_\rho^2\Big]
 + \sum_{\rho=3}^4 \Big[{\cal M}_k(i_B=2)(r^2 
S_\rho^2- C_\rho^2)+ \Gamma_\rho^2\Big]~, \\
D_r &=& {\cal W}_{p,k}\Big[\Gamma^2+r^2{\cal M}_k^2(i_B=2)\Big]~.
\end{eqnarray}
Fourth branch: $ap_\mu~=~(\pi,~\pi,~\pi,~0)$ 
or three other permutations. Explicitly 
for $ap_\mu~=~(\pi,~\pi,~\pi,~0)$  
\begin{eqnarray}
N_r &=&- \sum_{\rho=1}^3 \Big[{\cal M}_k(i_B=3)
(S_\rho^2-r^2 C_\rho^2)+ \Gamma_\rho^2\Big]
 + \Big[{\cal M}_k(i_B=3)(r^2 
S_\rho^2- C_\rho^2)+ \Gamma_\rho^2\Big]_{\rho=4}~, \\
D_r &=& {\cal W}_{p,k}\Big[\Gamma^2+r^2{\cal M}_k^2(i_B=3)\Big]~.
\end{eqnarray}
Fifth branch: $ap_\mu~=~(\pi,~\pi,~\pi,~\pi)$. 
\begin{eqnarray}
N_r &=& -\sum_{\rho=1}^4 \Big[{\cal M}_k(i_B=4)
(S_\rho^2- r^2 C_\rho^2)+ \Gamma_\rho^2\Big]~,\\
D_r &=& {\cal W}_{p,k}\Big[\Gamma^2+r^2{\cal M}_k^2(i_B=4)\Big]~.
\end{eqnarray}

\begin{table}
\begin{center}
\begin{tabular}{|l|l|l|}
\hline \hline
Branch &\multicolumn{2}{c|}{$\delta~m/(g^2~C_F)$}\\
\cline{2-3}
&  {Sunset} & {Tadpole} \\
\hline
{First}&{-0.0158} &{-0.3099} \\
\hline
{Second}&{+0.0148} &{-0.1549} \\
\hline
{Third}&{0.0000} &{0.0000} \\
\hline
{Fourth}&{-0.0148} &{+0.1549} \\
\hline
{Fifth}&{+0.0158}&{+0.3099} \\
\hline\hline
\end{tabular}
\caption{Numerical values of the additive mass shift for fermion at different branches for $r=1$ and $L=200$.}
\label{nuself}
\end{center}
\end{table}
\begin{figure}
\begin{center}
 \includegraphics[width=3.5in,clip]
{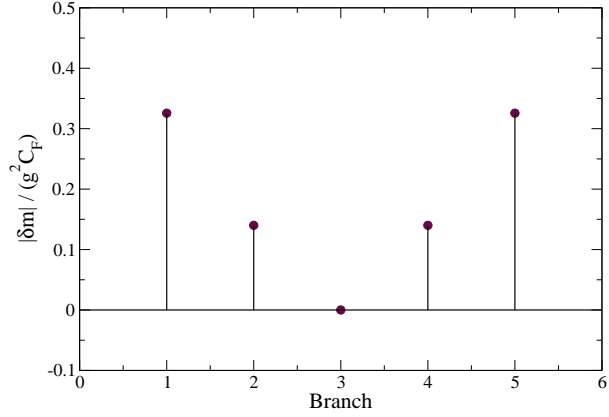}
\caption{The magnitude of the total additive mass shift (tadpole + sunset) plotted
versus the branch number.}
\label{shift}
\end{center}
\end{figure}
In table \ref{nuself} we present the numerical values of the additive mass shift 
separately from sunset and tadpole contributions 
for the fermion at different branches.
In figure \ref{shift} we plot the magnitude of the total additive mass shift 
(tadpole + sunset) versus the branch number. Note that the shift is maximum for 
the first and the fifth branches and is minimum (zero) for the third (central) branch.
The absence of additive renormalization in fermion self energy 
for the central branch to ${\cal O}(g^2)$ is explicitly shown also in ref. \cite{misumi}.

\section{Chiral Anomaly}
Now we consider the flavor singlet axial Ward Identity 
\begin{eqnarray}
\langle {\Delta}^{b}_{\mu} J_{5 \mu}(x) \rangle = 2m \langle
{\overline  \psi}_x \gamma_5 \psi_x \rangle + \langle \chi_x\rangle
\end{eqnarray}
where $\langle  {\cal O}  \rangle $ denotes the functional
average of ${\cal O}$. Explanation of
other terms are as follows:\\ 
\begin{eqnarray}
\Delta^b_\mu f(x) &=& \frac{1}{a} \left  [ f(x) - f(x -\mu) \right ]~,~
{\rm the~ backward~ derivative}, \\
J_{5 \mu}(x) &=& \frac{1}{2} \left [
{\overline \psi}_x \gamma_\mu \gamma_5 U_{x,\mu} \psi_{x+\mu}
+ {\overline \psi}_{x+\mu} \gamma_\mu \gamma_5 U^\dagger_{x \mu}
\psi_x \right ],~ {\rm the~axial~vector~current~} 
\\{\rm and}~~ 
\langle \chi_x \rangle &=& 2 ~ g^2 ~ \epsilon_{\mu \nu \rho \lambda} ~
{\rm Trace} ~F_{\mu \nu}(x)~ F_{\rho \lambda}(x)~ \frac{1}{(2 \pi)^4}\sum_p 
~\cos(p_{\mu}a)~{\cos}(p_{\nu}a)~ {\cos}(p_{\rho}a)~ \nonumber \\
& {\hspace{.3in}}\times r & {\cal M}_p(i_B) 
~\Big [ {\cos}(p_\lambda a)~ [a m + r {\cal M}_p(i_B)]
                             - ~ 4 r ~{\sin}^2 (p_\lambda a) \Big ]
({\cal G}_p(i_B))^3 , \label{caw1}  \nonumber \\
&=& - \frac{g^2}{16 \pi^2} ~ \epsilon_{\mu \nu \rho \lambda}~
{\rm Trace} ~F_{\mu \nu}(x)~ F_{\rho \lambda}(x)~ I(am,r,L). \label{caw2}
\end{eqnarray}
Here 
\begin{eqnarray}
{\cal G}_p(i_B) &=& \Bigg (
\sum_\mu {\sin}^2 (a p_\mu ) +  \Big [a m + r {\cal M}_p(i_B)\Big]^2 \Bigg )^{-1}.
\end{eqnarray}
Explicitly, $ \sum_p = (\frac{2 \pi}{L})^4 \sum_{n_1,n_2,n_3,n_4}$.
  In all our plots it is 
the anomaly integral denoted by the function $I(am,r,L)$ which we have plotted.

Following Karsten and  Smit \cite{ks}, the 
limits on the momentum sum are changed from 
$ (- \pi, + \pi) $ to 
$ (- \pi/2,3\pi/2) $ and further  the momentum 
sum 
hypercube is divided into 16 smaller hypercubes corresponding to  
$(- \pi/2, +\pi/2) $ and $(+ \pi/2, + 3 \pi/2)$
for each  $a p_\mu, \mu=1,2,3,4 $. Thus the total anomaly contribution is 
decomposed into the contributions from five different types of species and the anomaly 
integral takes the form  $I=I_0-4I_1+6I_2-4I_3+I_4$.
In $I_0$ all the four momentum integrations
range from $(- \pi/2, +\pi/2) $ and in $I_4$ they range from 
$(+ \pi/2, + 3 \pi/2)$. In  $I_1$ one of the momentum integrations range
from  $(+ \pi/2, + 3 \pi/2)$, the rest from $(- \pi/2, +\pi/2) $
and vice-versa for $I_3$. In the third (central) branch $I_2$ two momentum
integrations range from $(+ \pi/2, + 3 \pi/2)$ and the rest from 
$(- \pi/2, +\pi/2) $.

\begin{figure}
\subfigure{   
 \includegraphics[width=3.5in,clip]
{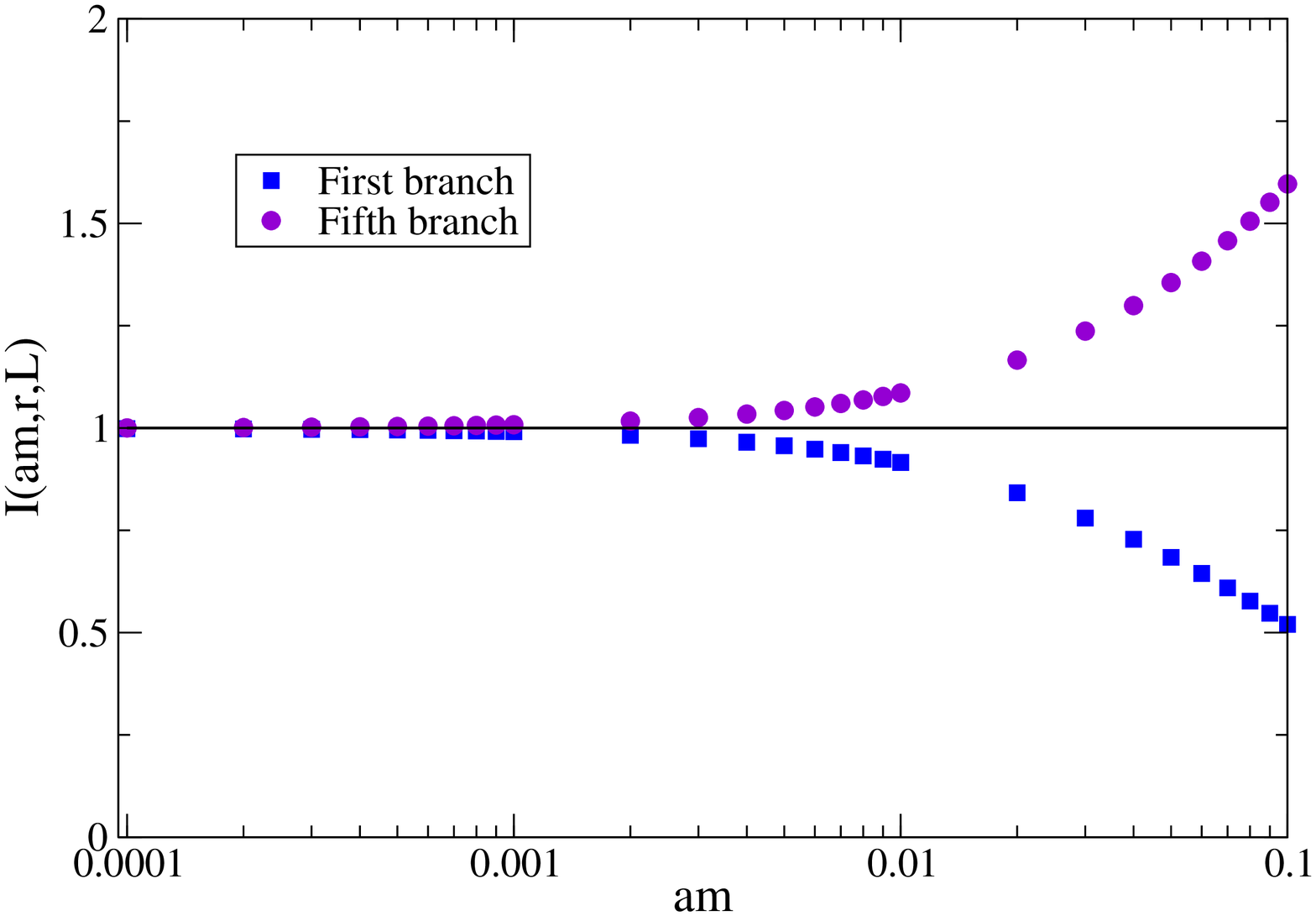}}
\subfigure{
 \includegraphics[width=3.5in,clip]
{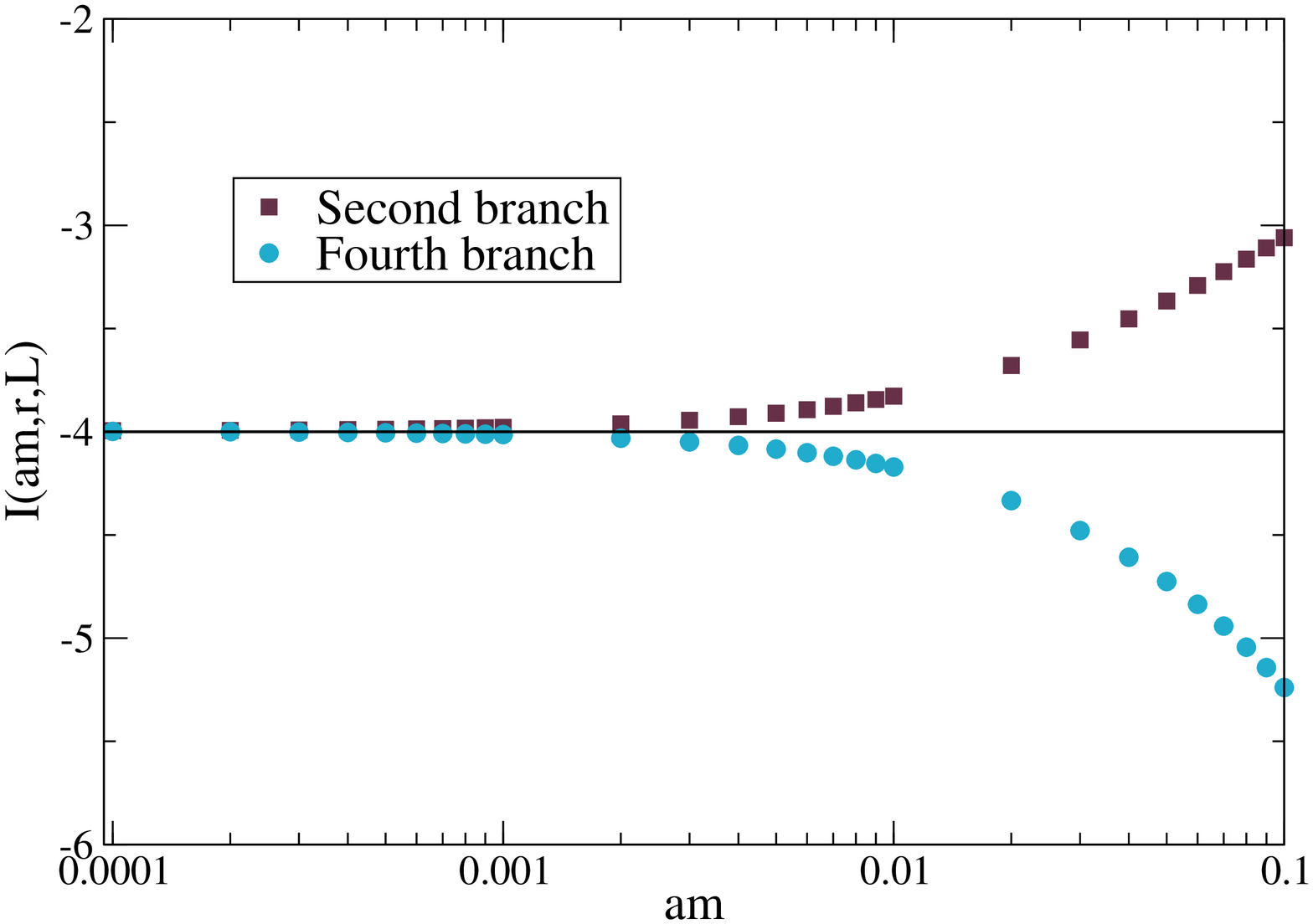}}
\caption{The function $I(am,r,L)$ for $r=1.0$ and
$L=100$ as a function of $am$ for the first and fifth branches (left) and for 
the second and fourth branches (right).}
\label{noncentral}
\end{figure}

\begin{figure}
\begin{center}
 \includegraphics[width=3.5in,clip]
{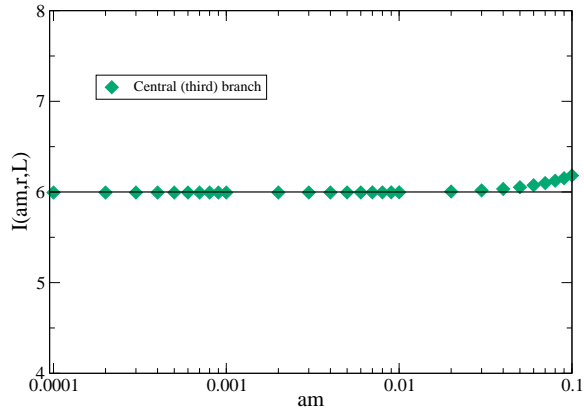}
\caption{The function $I(am,r,L)$ for the central branch for $r=1.0$ and
$L=100$ as a function of $am$.}
\label{central}
\end{center}
\end{figure}

First, to perform the integration analytically, we set the bare mass $am=0$,
use the identity \cite{ks}
\begin{eqnarray}
\Big [{\cal M}_p(i_B)\Big]^2~ \cos (ap_\beta) ~-~4~r~{\cal M}_p(i_B) ~ 
\sin^2(ap_\beta)  ~&=&~\Bigg [ \Big [{\cal M}_p(i_B)\Big]^2 + \sum_\sigma 
\sin^2(ap_\sigma) \Bigg]^3 ~~ \times\nonumber \\
 &&\frac{\partial}{\partial(ap_\beta)} ~
\Bigg [ \sin (ap_\beta)  \Bigg\{  \Big [{\cal M}_p(i_B)\Big]^2 + \sum_\sigma 
\sin^2(ap_\sigma)\Bigg\} \Bigg]^{-2}~    \nonumber \\
\end{eqnarray}
and do a partial integration. In the infinite volume continuum limit, the results
for the integrals are as follows.\\
First branch: $I_0 \rightarrow 1$, $~I_1,~I_2,~I_3,~I_4~
\rightarrow 0$.\\
Second branch: $I_1 \rightarrow 1$, $~I_0,~I_2,~I_3,~I_4~
\rightarrow 0$.\\
Third (central) branch: $I_2 \rightarrow 1$, $~I_0,~I_1,~I_3,~I_4~
\rightarrow 0$.\\
Fourth branch: $I_3 \rightarrow 1$, $~I_0,~I_1,~I_2,~I_4~
\rightarrow 0$.\\
Fifth branch: $I_4 \rightarrow 1$, $~I_0,~I_1,~I_2,~I_3~
\rightarrow 0$.\\
Since numerical simulations are performed at finite volume and finite lattice 
spacing, it is of interest to study the effect of  symmetry violation on the 
anomaly integral as a function of the lattice fermion mass at finite volume \cite{anomaly1, anomaly2}.
In order to avoid the zero mode problem
we have used antiperiodic boundary condition in all four directions.
In figure \ref{noncentral} we plot 
the function $I(am,r,L)$ for $r=1.0$ and $L=100$ as a function of $am$ 
for the first and fifth branches (left) and for the second and 
fourth branches (right). In figure \ref{central}, we plot the function 
$I(am,r,L)$ for $r=1.0$ and $L=100$ as a function of $am$ for the central branch.
From figure 
\ref{noncentral} (left), we observe that the cut-off effects are almost equal and 
opposite for first and fifth branches. Similar observation can be made 
regarding second and fourth branches from figure \ref{noncentral} (right).
Comparing figures \ref{noncentral} and \ref{central}, we conclude that 
cut-off effects are minimal for the central branch. We have picked $L=100$ for our
plots as we have found that finite  
volume effects are negligible at this volume for the range of $am$ shown in the
figures.


\section{Discussion and Conclusions}
It is well known that the naive discretization of the fermionic action gives rise to 
sixteen degenerate species including  the desired physical one.
These sixteen species are grouped into five branches
with degeneracy (chirality) given by 1(1), 4(-1), 6(1), 4(-1) and 1(1), 
rendering the theory free of chiral anomaly. 
With the conventional Wilson term in the
continuum limit, apart from the first branch, species corresponding to all other
branches become infinitely massive and decouple from the theory thereby 
reproducing the correct chiral anomaly.
The branches other than the first one are rarely 
explored. However, recently the existence of an  additional symmetry 
in the central branch which prohibits additive renormalization of 
fermion mass has been discovered in the ref. \cite{kimura}.

In this work, in order to explore all branches we introduce a generalized 
Wilson term containing a branch selector index ($i_B$). By choosing $i_B$
one can make the fermions belonging to a particular branch physical. 
The fermions belonging to the rest of the branches become infinitely 
massive and decouple from the theory in the continuum limit. 
The conventional Wilson term corresponds
to $i_B=0$. To investigate the effect of radiative corrections, we calculate
the additive mass renormalization in fermion self-energy and the chiral anomaly
to ${\cal O}(g^2)$ in perturbation theory for all the branches.

First we summarize the results of additive mass shift from tadpole 
and sunset contributions. The tadpole contributions 
for the first and fifth branches are equal in magnitude but opposite in
sign. Same is true for the sunset contributions also. Thus $\delta m$ vanishes
if we average over the first and fifth branches. Similar observations hold
for the second and the fourth branches also. Coming to the central branch
the additive mass shifts from tadpole and sunset contributions separately 
vanish. This leads to the absence of additive mass renormalization in 
accordance with theoretical expectation.
In the calculation of chiral anomaly first we perform an analytical calculation
setting $am=0$ and using the Karsten-Smit identity. We find the correct value 
of the anomaly for different branches with corresponding degeneracy factors
and signs dictated by the chiral charges in the continuum limit.
Since numerical simulations are performed at finite volume, finite lattice spacing and
finite fermion mass, we have studied the effect of symmetry violation (given in 
eq. \ref{symmetry}) on the anomaly integral as a function of the lattice fermion mass.
The cut-off effects are almost equal in magnitude but opposite in sign for the first 
and the fifth branches. Same holds for the second and the fourth branches also.
The cut-off effect is minimal for the central branch.

In conclusion, our exploration of the different branches of the fermion doublers 
in perturbation theory, in the context of additive mass renormalization and chiral 
anomaly, has shown that by appropriately averaging over suitably
chosen branches one can reduce cut-off artifacts. Comparing the central branch with
all other branches, we find that the central branch, among all the avatars 
of the Wilson fermion, is the most suitable candidate
for exploring near conformal lattice field theories.



\end{document}